\documentstyle[prl,aps,epsfig,multicol]{revtex}
\begin{document}

\def\LCO{La$_2$CuO$_{4+\delta} $}
\def\LSCO{La$_{2-x}$Sr$_x$CuO$_{4} $}
\def\LNSCO{La$_{1.6-x}$Nd$_{0.4}$Sr$_x$CuO$_{4} $}
\def\YBCO{YBa$_2$Cu$_3$O$_{7-\delta}$}
\def\BSCCO{Bi$_2$Sr$_2$CaCu$_2$O$_{8+\delta}$}
\def\be{\begin{equation}}
\def\ee{\end{equation}}
\def\ba{\begin{eqnarray}}
\def\ea{\end{eqnarray}}
\twocolumn[
\hsize\textwidth\columnwidth\hsize\csname@twocolumnfalse\endcsname
%\draft

\title{Thermodynamics of the interplay between magnetism and
high-temperature superconductivity}

\author{ Steven A. Kivelson$^1$, G. Aeppli$^2$,and Victor J. Emery$^3$}
\address{$^1$Department of Physics and Astronomy, University of
California
Los Angeles\\ Los Angeles, California 90095-1547}
\address{$^2$NEC Research Institute, 4 Independence Way, Princeton, NJ
08540}
\address{$^3$Brookhaven National Laboratory,Upton NY 11973}
\date{\today}
\maketitle

\begin{abstract}
The copper-oxide based high temperature
superconductors have complex phase diagrams with multiple ordered phases. It even
appears that the highest superconducting transition temperatures for certain
cuprates  are found in samples which display simultaneous onset of magnetism and
superconductivity. We show here how the thermodynamics of fluid mixtures - a
touchstone for chemistry as well as hard and soft condensed matter physics - accounts
for this startling observation, as well as many other properties of the cuprates
in the vicinity of the instability towards ``striped'' magnetism.
\end{abstract}
%\pacs{PACS:}
\bigskip
]

The phase diagrams of conventional superconductors are usually 
simple,
with no ordered phases competing with the superconducting state. By
contrast, the high temperature
superconductors
have a number of
competing phases which appear as the temperature is lowered.
One of the most astonishing manifestations of this competition occurs in
the
LaCuO family of materials  
where coexisting magnetism and high temperature superconductivity 
\cite{tranquada1,tranq2,ichikawa,niedermayer,nachumi,nmr,savici,yamada,lee}
has
been reported. 
In La$_{1.6-x}$Nd$_{0.4}$Sr$_x$CuO$_4$, long period magnetic
(``stripe'') order
(as 
detected by neutron diffraction) sets in at a higher temperature
than the 
superconducting $T_c$ (and indeed, charge ``stripe'' order appears at a
still higher temperature\cite{tranquada1,ichikawa}).  
In La$_2$CuO$_{4.12}$ and La$_{1.88}$Sr$_{0.12}$CuO$_4$,
superconductivity and 
long-period magnetism appear to have the same onset temperature in the 
same bulk crystal!  However, while muon spin relaxation data tells a 
grossly similar story\cite{niedermayer,nachumi} in 
La$_{1.6-x}$Nd$_{0.4}$Sr$_x$CuO$_4$, the corresponding data\cite{savici}
for La$_2$CuO$_{4.12}$  show that the
magnetism is peculiar in that it resides in only a fraction of the
sample, and its temperature evolution is due to growth of the magnetic 
fraction rather than an increase in the order within the magnetic 
fraction. We show how these observations can be understood from the
classical thermodynamics of
two-phase  mixtures, which is applicable because of the well-documented
tendency of 
antiferromagnets to expel holes\cite{russian,ek,other}. 

{\bf Phase Diagrams with Competing Orders:  }
The interplay between ``stripe'' magnetism and
superconductivity can be understood most simply
by treating the liquids of mobile
charge carriers in the high-temperature superconductors as fluids with a
variety of ground states. As for other complex fluids,
%such as liquid crystals, 
the coupling between the order parameters 
%forthe various ground states 
can lead to phases with mesoscopic density
modulations as well as diverse combinations of the order parameters
themselves. To make this phenomenon explicit, we follow a standard
paradigm of statistical physics and consider\cite{blount} the simplest
Landau free energy, $F$, for two coupled order parameters,
$\vec S$ and $\Delta$, which represent the long
period antiferromagnet and the superconducting order 
%parameters 
respectively.  Spin rotation invariance and gauge
invariance ($\Delta$ is a complex number whose phase cannot influence
the free energy) imply that $F$ 
is a function of $|\vec S\cdot \vec S|$ and $|\Delta|^2$:
\ba
F=&& F_0(\mu,T)+\alpha(\mu,T) |\vec S\cdot \vec S| + \beta(\mu,T) |\vec
S\cdot \vec S|^2  \\
\label{eq:F}
&&+a(\mu,T) |\Delta|^2 + b(\mu,T) |\Delta|^4 +\gamma(\mu,T) |\Delta|^2
|\vec S\cdot \vec S| ...\nonumber
\ea
where $T$ is  the temperature, 
$\mu$ is the chemical potential for doped holes, $...$ represents higher
order terms in 
powers of the ordering fields, and the various coefficients embody the 
effects of all the short-distance physics.  The phase diagram is then 
determined (at mean-field level) by minimizing $F$ with respect to $\vec
S$ and $\Delta$.  While such a mean-field description ignores important
fluctuations, especially given the fact that the high temperature
superconductors are quasi-two dimensional, it provides a valid zeroth
order
way to examine the global structure of the phase diagram.

{From} the macroscopic viewpoint adopted in the present paper, the
parameters which enter
the Landau free energy in Eq. (1) are purely phenomenological.  However,
some
insight 
concerning the microscopic physics can be inferred from the behavior of
these
parameters.  In particular,  if $\gamma>0$,
superconductivity and long period magnetism
compete, while if $\gamma < 0$, they enhance each other.  Indeed,
Tranquada and
collaborators\cite{ichikawa} have concluded that static magnetism and
superconductivity compete, and this is
certainly intuitively sensible. Recent experiments on
the behavior of vortex cores\cite{new1} and on
superconductivity-induced changes in the magnetic
susceptibility\cite{lakegap} of optimally doped {\LSCO} confirm
that $\gamma >0$,
in agreement with these arguments. One main purpose of this paper is to
show how even if $\gamma >0$, magnetism and superconductivity can 
%show
set in at the same temperature in a single sample. 

An important subtlety arises from the fact that, as in many experiments
on classical fluids, it is the total number of constituents of the fluid
rather than the chemical potential that is fixed. For the cuprates, the
constituents are the charge carriers (doped holes) and their number is
fixed by 
the chemical composition of the compound under study ({\it e.g.} the $x$
in
{\LSCO}).  Therefore
$\mu$ must be determined from the implicit relation
\begin{equation}
-x=\partial F/\partial \mu=F_0^{\prime} + {\alpha}^{\prime}|\vec S\cdot
\vec S|+a^{\prime} |\Delta|^2
+...
\label{eq:mu}
\end{equation}
where $x$ is the concentration of doped holes, $\prime$ denotes
differentiation with respect to $\mu$,
and
$\vec S$ and $\Delta$ are the
equilibrium values of the ordering fields as a function of $T$ and
$\mu$.  However, where two-phase
coexistence occurs in the phase diagram there are values of
$\mu$  at which $\partial F/\partial \mu$ has a discontinuity.  In this
case, Eq. (\ref{eq:mu}) 
has solutions for fixed $\mu$ at two different values of $x$, $x_1$ and
$x_2$.
% but has no solutions for $x_1 < x < x_2$. 
The equilibrium state for fixed $x$ in
this range consists of
a two phase mixture, with the volume fractions  of the hole rich
($x=x_2$) and hole poor
($x=x_1$) phases determined by the classical lever rule,
$f_1=(x_2-x)/(x_2-x_1)$, and $f_2=1-f_1$.
Otherwise, in those ranges of $x$ for which the equilibrium state is
single phase, it is possible,
if desired, to perform a Legendre transform, so that the coefficients in
the Landau free energy
($\alpha$, $\beta$, etc.) are expressed as functions of $x$ and $T$.

\begin{figure}
\begin{center}
%\leavevmode
\noindent
\epsfxsize=3in
\epsfbox{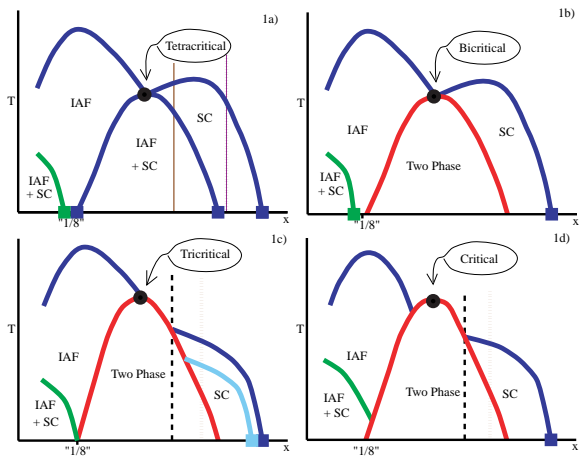}
\end{center}
%\noindent
%\centerline{\includegraphics[width=0.6\columnwidth,angle=0]{aeppli.eps}}
\caption{Schematic phase diagrams derived from the Landau free energy
under the various
conditions described in the text.  ``IAF'' and ``SC'' indicate
incommensurate (striped)
antiferromagnetic  and superconducting order, respectively.  The 
circles represent classical
critical or multicritical points
and the squares quantum
critical points.  The various vertical lines represent trajectories
through the
phase diagram discussed in the text.  The pale blue phase boundary 
in c) represents the effect of an applied magnetic field on the
phase diagram.}
\end{figure}

Generally, broken symmetry phases occur only at lower temperatures, so
it is
reasonable to expect  $\alpha$ and $a$ to  change
sign at bare transition temperatures defined according to
$\alpha(x,T)\equiv
\alpha_0(x,T)[T-T_s(x)]$ and
$a(x,T)\equiv  a_0(x,T)[T-T_{\Delta}(x)]$ where $\alpha_0$, $a_0 >0$.  
(Here, we use the same symbol for the original and Legendre transformed
coefficients in the Landau
theory.)  On both theoretical
and empirical grounds we expect the native magnetic ordering
temperature 
$T_s(x)$ to be a generally decreasing function of $x$, reflecting the
frustration
of  hole-motion in the magnetically ordered state.  However, a
peak in $T_s(x)$ at a special ``commensurate'' value of $x$ can occur
when the period of the stripe order is a small integer times the
underlying
crystalline lattice constant;  such a commensurability
effect gives rise\cite{tranquada1,ichikawa} to the
anomalous stability of the long period magnetism at
$x=1/8$, as shown schematically in Fig. 1.
$T_{\Delta}(x)$ is 
known empirically to be a non-monotonic function of $x$, reaching a peak
around an optimal value of $x\sim 0.15$ and dropping slowly as $x$ is 
increased or decreased.

There are several generic phase diagrams derivable from 
such a Landau free energy.  Most simply, a tetracritical point can occur
at $(T, \mu)=(T^*,  \mu^*)$ 
which leads to  a phase diagram of the type shown in Fig. 1a.  Here, in
addition to the 
phases with either magnetic or superconducting order, there is a
homogeneous
intermediate phase with  bulk coexistence of the two orders.  
We have included in this and all
subsequent frames of Fig.1 a second superconducting phase ( which is
permitted, but not
required in the simplest Landau theory) at hole concentrations $x <
1/8$.
This is motivated by the observation that many of the high temperature
superconductors exhibit multiple humps
in the superconducting transition temperature plotted against hole
concentration, most
notably the {\YBCO} (especially when lightly Zn doped\cite{zndoped}) and
{\LSCO}
related compounds. As mentioned above, we have also indicated a peak in
the
magnetic ordering temperature at $x=1/8$.  The corresponding
anomalous supression of the superconducting transition temperature
at this point shown in the figure
is a consequence\cite{ichikawa} of the peak in the magnetic ordering
under the
assumption that $\gamma >0$, even if $T_{\Delta}(x)$ is monotonic in
this range of
$x$. 

The tetracritical point obtains as long as solutions exist to the
simultaneous equations
$\alpha(\mu^*,T^*)=0$ and $ a(\mu^*,T^*)=0$ which at the same time
satisfy
the inequalities $
\beta(\mu^*,T^*) > 0$, $ b(\mu^*,T^*) > 0$, 
and $ \beta(\mu^*,T^*)  
b(\mu^*,T^*) > 4  \gamma(\mu^*,T^*)^2$. If, on the other
hand, the last of these inequalities is violated, ${\it 
i.e.}$ if $ \sqrt{\beta(\mu^*,T^*)  b(\mu^*,T^*)} < 2  
\gamma(\mu^*,T^*)$, the tetracritical point is replaced by a bicritical
point.  
Not only does that mean that there is no
phase with bulk 
coexistence of the two orders, it also means that below the bicritical 
point, there is a region of two-phase coexistence, where a hole-poor, 
magnetically ordered and a hole-rich, superconducting state coexist, as
shown in Fig. 1b.  In reality, this coexistence must not be taken 
literally.  Because of the long-range Coulomb interaction between holes,
macroscopic phase separation is thermodynamically forbidden.  Where 
macroscopic phase coexistence would occur in a neutral system, a form of
Coulomb frustrated phase separation\cite{ek} is expected, leading to a 
state which is inhomogeneous on an intermediate length scale.  This also
means that the two, coexisting phases are in microscopic proximity to
each 
other, and hence that a modicum of superconducting order will be induced
in the magnetic regions, via the proximity effect, and
conversely\cite{ak}. Such competition is a recurring theme in this
problem;
there are
empirical and theoretical reasons to  believe\cite{PNAS,zaanen} that the
long
period antiferromagnetic stripe
order,  itself, is at least in part a consequence of Coulomb frustrated
electronic 
phase separation on a smaller length scale.

A third possibility shown in Fig. 1c 
occurs
if there is a tricritical point, where
$ \alpha(\mu^*,T^*)= \beta(\mu^*,T^*)=0$, while all the other 
coefficients (including certain higher order terms, not discussed 
explicitly) remain positive.  This leads to a phase diagram of the sort
shown in Fig. 1c.  Here, superconductivity
%, if it appears at all,
manifests itself
below a phase-boundary that terminates on the edge of the two-phase
region 
in a critical end point.

For completeness, we present a fourth possible phase diagram topology,
shown in Fig. 1d, for which a more thorough analysis of the 
free energy function is necessary.  Here, instead of a multicritical 
point, we consider the occurrence of a simple critical point, below
which  phase separation occurs into hole rich and hole poor phases,
neither of
which is ordered.  In this case, both the antiferromagnetic and the 
superconducting phase boundaries terminate at critical end points.  

{\bf  Relation to Experiment in {\LCO}:  }
The thermal
evolution of a given material should be associated with a trajectory in
one of the generic phase diagrams in Fig. 1. In particular, we propose
associating {\LCO} with
the vertical dashed line in Fig. 1c or 1d.  It is a special trajectory,
in the
sense that it is tuned to pass close to the critical end point, but this
requires fine tuning of only one parameter.  Below $T_c$, the system
forms  an inhomogeneous mixture of a high density superconducting and a
low  density antiferromagnetic phase.
At T$_c$, the sample is a single-phase
superconductor, with a superconducting volume fraction
$f_{SC}(T)=1-f_{Mag}(T)$ that shrinks at the
expense of  the antiferromagnet
as $T$ is reduced through the two-phase region. Since there are no
critical 
effects on the shape of the phase boundary associated with a critical
end
point,
just below $T_c$, $f_{Mag} = A(T_c-T) + \ldots$, where $A$ is determined
by the slope
of the phase boundary.  Since the magnetic
ordering  of the hole poor phase would set in at a temperature well
above $T_c$,
the  ordered moment $M(T)$ in the antiferromagnetic fraction is
immediately 
large, and essentially temperature independent!  The
growth of the magnetism is associated more with the growth of the hole-poor
fraction rather than with the rise of the order parameter within the
hole-poor regions.  

Not only is this scenario consistent with the
simultaneous onset of superconductivity and magnetism, it also
reconciles the neutron
scattering and
$\mu$-SR data, which
we reproduce in Fig. 2.  Specifically, in a two-phase mixture in which
only
one phase is magnetic, the intensity $I$ of the Bragg scattering
measured by neutrons is related to $f_{Mag}$ and
$M$, 
measured in $\mu$-SR, according to
the relation
\begin{equation}
I(T)=M^2(T) f_{Mag}(T).
\label{eq:I}
\end{equation} 
\begin{figure}
\begin{center}
%\leavevmode
\noindent
\epsfxsize=3in
\epsfbox{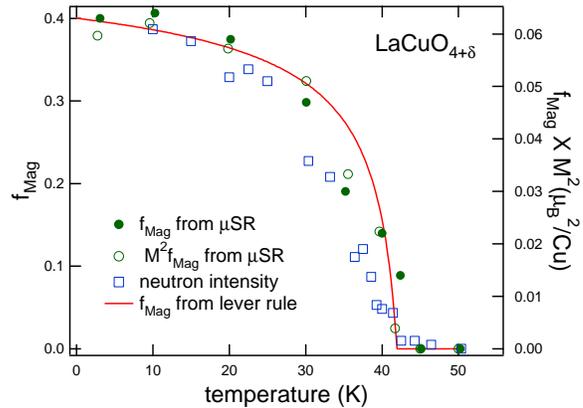}
\end{center}
%\noindent
%\centerline{\includegraphics[width=0.6\columnwidth,angle=0]{aeppli.eps}}
\caption{Interplay between magnetic and superconducting order in {\LCO}.
The open and closed circles represent, respectively the antiferromagnetic
fraction, $f_{Mag}(T)$, and the product $M^2(T)f_{Mag}(T)$  from the
$\mu$-SR data in Ref. {\protect \ref{savici} }.  The open squares
represent the neutron intensity $I(T)$ from Ref. {\protect \ref{lee} }, scaled by a
factor of 2.7.  The solid line is a theoretical
prediction for $F_{Mag}(T)$, using the lever rule,
for the vertical dashed trajectory in the tricitical phase diagram in
Fig. 1c,
assuming the two-phase region is bounded by the curve $T_{2}= 4T_0
(x_1-x)(x-x_2)/(x_2-x_1)^2$ with parameters (discussed in the text) $T_0=43.7$K,
$x_1=0.125$,
$x_2=0.188$, and with a mean hole density $\bar x=0.15$, 
representative{\protect\cite{lee}} of stage IV {\LCO}. }
\end{figure}

The fact that the absolute neutron
intensity\cite{lee} at low temperature is 0.022 $\mu_B^2$, which is 40\%
of that anticipated from the muon experiments according to Eq. \ref{eq:I} suggests
that either the sample interior penetrated by the neutrons is different from the
surface region probed by the muons, or that the integration of the magnetic
signal in the neutron experiment could be incomplete. 
%(We look forward to
%future muon and neutron experiments on the same samples to resolve the
%origin of the current discrepancies.)
In the figure, we have scaled the neutron data so that it matches 
the inferred intensity from $\mu$-SR data at $T=0$;
the (scaled)
neutron data still falls somewhat short of
$M^2(T) f_{Mag}(T)$ immediately below T$_c$, 
%probably
possibly because the neutron data are peak
intensities, rather than integrals over the three dimensional (in
reciprocal space) structures containing the net spectral weight 
responsible for the muon data. More specifically, if there is any
broadening in the peaks as T$_c$ is approached, the peak intensity will
be reduced relative to the integral. 
The red line in the figure corresponds to the lever rule prediction for
an average hole density\cite{lee} $\bar x=0.15$ and 
assuming the simplest possible parabolic form for the bounding curve
of the 2-phase region
$T_2(x)=4T_0 (x_2-x)(x-x_1)/(x_2-x_1)^2$.  So as to have no free parameters
in determining the theoretical curve, we have taken
$x_1=1/8$, reflecting the special 
stability of the striped phases at $x\approx 1/8$.   The remaining parameters can be
deduced directly from
$f_0\equiv f_{Mag}(0)=0.4$  
and the condition that antiferromagnetism and superconductivity onset at the
same temperature, $T_2(\bar x)=T_c=42$K, according to
$T_0=T_c/[4f_0(1-f_0)]$ and
$x_2-x_1=(\bar x - x_1)/f_0$.  It turns out that the theoretical
curves are very sensitive to the exact value of $f_0$.  For instance, the
quality of the fit can be improved if we take $f_0=0.35$ which is somewhat smaller
than the reported value, although possibly within experimental uncertainty.   

In the discussion above, we have ignored non-electronic physics. The
most obvious possibility here is that motion of excess oxygen results in
an inhomogeneous distribution of oxygen, which in turn would lead to
inhomogeneous hole density and electronic properties. The final outcome
would
be an inhomogeneous distribution of magnetism and superconductivity,
even while charge neutrality would obviously be satisfied on a local
scale. Nonetheless, our sense is that this explanation of the data is
improbable. First, the onsets of superconductivity and magnetism
coincide\cite{yamada} in {\LSCO} as well as {\LCO} in the same range of
average hole density. Second, the findings of Lee {\it et al}\cite{lee}
provide an
important clue concerning how charge neutrality is preserved at long
length scales without
substantial oxygen motion;  they report magnetic order with a remarkably
long correlation length (greater than $125\AA$) in the basal plane, 
but with  interplanar correlations extending
only over 2-3 planes. Charge neutrality can therefore be preserved over
distances of order 10-15 Angstroms, even while the system breaks up into
thin magnetic and superconducting layers - ``pancakes" parallel to the
basal planes. Finally, NMR studies\cite{imai} suggest that the material
is
electronically
single-phase at temperatures above $T_c$, implying that the observed
inhomogeneities are induced by the onset of order.

{\bf {\LSCO}, {\LNSCO}, etc.:  }
The phase diagrams in Fig.1
provide a framework for understanding many other properties of the
lanthanum cuprate family. 
To begin with, a miscibility gap leading to coexistence of
superconducting and
non-superconducting phases readily accounts for the finding of optimal
Meissner fractions
for {\LSCO} only near special hole densities\cite{harshman}. In
addition,
as for the superoxygenated {\LCO} discussed above, the ordered magnetic
moments
deduced from neutron diffraction\cite{yamada} are less than the frozen
local
moments deduced from muon spin relaxation\cite{niedermayer} and much
less than that
seen for ordinary insulating two-dimensional antiferromagnets, implying
also that the
magnetic order resides in only a part of the sample. The appropriate
phase diagram
for {\LSCO} might then look like a disorder broadened (glassy) image of
Fig. 1b. On
the other hand, as Nd is inserted, the magnetism (as detected in neutron
diffraction)
becomes stronger\cite{tranquada1,tranq2,ichikawa} and seems to appear
throughout the
sample volumes\cite{nachumi}, even while superconductivity survives.
There is also
remarkable evidence for a non-monotonic temperature dependence of the
superfluid
density\cite{tajima}, which implies that on cooling,
La$_{1.55}$Nd$_{0.3}$Sr$_{0.15}$CuO$_4$ first undergoes a transition to
a uniform
superconducting state, and then to a state with coexistence of magnetism
and
superconductivity.  Thus, the tetracritical diagram, Fig.1a might be
more appropriate
for {\LNSCO}, with a microscopic coexistence of magnetic and
superconducting order;
La$_{1.55}$Nd$_{0.3}$Sr$_{0.15}$CuO$_4$ might then be represented by the
solid brown
trajectory in that figure. Finally, La$_{1-x}$Ba$_x$CuO$_4$ exhibits two
separated
superconducting ``domes'', with an intervening magnetic regime (at
$x=1/8$) which is
magnetic and not superconducting\cite{lbco,axe,luke,kumagai}. The
corresponding phase
diagram could therefore be that shown in Fig. 1a or 1b. Of course, any
real material
exists not on a one-dimensional axis representing the doping, but rather
in a
multidimensional space spanned by the parameters required to shift from
diagram to
diagram in Fig. 1. The outcome is then that, as demonstrated by
inelastic neutron
scattering from La$_{1.86}$Sr$_{0.14}$CuO$_4$\cite{aeppli}, optimally
doped {\LSCO},
described by the dotted trajectory in Fig 1a, can show behavior
associated with the quantum critical point where magnetism disappears in
the tetratcritical diagram. 

{\bf Additional Details:  }
With the
exception of {\LCO}, all the materials discussed have intrinsic disorder
due to the
random arrangement of the dopant atoms;  any stripe ordering transition
is thus
expected\cite{rome1} and observed\cite{tranqglassy} to be intrinsically
glassy, with
the ordered phase being a ``stripe glass.'' It should also be clear that
the interplay
between magnetism and superconductivity, 
which accounts for so many 
key observations, omits other features\cite{kfe,sudip,chandra,subir}
expected or observed, of the actual phase diagrams.
Within the magnetic regime
there can be a variety of phases which we have not indicated:
the long period magnetic order
can be commensurate with the underlying lattice or incommensurate, and
the stripes can point along the
copper oxide bonds (``horizontal'') or\cite{diag} at 45$^o$ to them
(``diagonal'').
Finally, we have not shown
the transitions involving charge order although certainly, at least for
{\LNSCO}, there
is a separate transition\cite{tranquada1,ichikawa} at which
unidirectional charge
density wave (``charge stripe'') order appears at a
temperature above the magnetic and superconducting transitions.  

{\bf Prospects:  } A miscibility gap in the phase diagram not only
resolves old puzzles, but also provides a framework for understanding
current and
future experiments.
Particularly important are measurements of magnetic field-dependent
effects, as they
permit a continuous variation of the parameters in the Landau free
energy.  
%Many predictions follow readily from the simple
%analysis presented here.  
For example, if {\LCO} in zero field happens
to lie on a trajectory which passes through a critical end point, as we
have supposed, then in
a magnetic field, which will suppress the superconducting
$T_c$, magnetic Bragg scattering, originating from a small
magnetic fraction, will still appear at a temperature roughly equal to
the
zero field $T_c$. This is illustrated in Fig. 1c, where the field
shrinks the superconducting region of the phase diagram from the solid to
the lighter blue line. However, as we pointed out above,
the vortex state is complicated -  it is clear that
physics beyond the simple Landau theory needs to be
invoked\cite{eugene} in order to understand the dramatic magnetic field-induced
increases in the
antiferromagnetic Bragg intensities.
% which are not anticipated from simply applying the
%lever rule (which would make them field-independent). 
This physics is clearly beyond
the scope of the present work, but may relate to early theory indicating
that vortices in superconductors derived from Mott-Hubbard insulators
are insulating nano-antiferromagnets, with a different charge density
than the surrounding
superconductor\cite{zhang}.
Magnetoresistance data\cite{andoh} indicate
enhanced insulating tendencies 
%in {\LSCO} 
for $x$ near 1/8, {\it i.e.}
fields above $H_{c2}$
uncover the behavior also seen when superconductivity is suppressed by
chemical pressure. Furthermore, neutron diffraction reveals
field-induced magnetic Bragg scattering, which sets in near the
zero-field critical temperature for {\em superconductivity}\cite{new2}.
Even for samples beyond the miscibility gap,  
because type II
superconductors below H$_{c2}$ are heterogeneous mixtures of ``normal''
vortices and superconducting material, the vortices can exhibit - on a
finite length scale - the magnetism one might have expected if
superconductivity had not intervened (assuming $\gamma >0$). Recent
inelastic neutron scattering
experiments\cite{new1} on optimally doped {\LSCO} 
are in agreement with this
expectation -  the vortices are found to behave as nanomagnets with
growing ``stripe'' order with
decreasing temperature.

\noindent{\bf Aknowledgements:}  We wish to acknowledge important
conversations with S.~Brown,
S.Chakravarty, L.Gorkov, Z.Fisk, T. Imai, A.Kapitulnik, C.Renner, T. Uemura, and J.
Tranquada, and technical assistance from S.~Brown. 
This work was initiated
while two of us (SAK and GA) were participants in the High T$_c$ Program
at ITP-UCSB.
SAK was supported, in part, by NSF grant number DMR98-08685 at UCLA.

\end{document}